\documentclass[twocolumn]{aastex631}

\usepackage{graphicx}
\usepackage{amssymb}

\received{}
\revised{}
\accepted{}
\submitjournal{AAS}

\shorttitle{EL CVn-type Eclipsing Binary J0247-25}
\shortauthors{Kim et al.}

\begin{document}

\title{Pulsation and Rotation of the EL CVn-type Eclipsing Binary 1SWASP J024743.37-251549.2}
\correspondingauthor{Seung-Lee Kim}
\email{slkim@kasi.re.kr}

\author[0000-0003-0562-5643]{Seung-Lee Kim}
\affil{Korea Astronomy and Space Science Institute, Daejeon 34055, Republic of Korea}

\author[0000-0002-5739-9804]{Jae Woo Lee}
\affil{Korea Astronomy and Space Science Institute, Daejeon 34055, Republic of Korea}

\author[0000-0003-0043-3925]{Chung-Uk Lee}
\affil{Korea Astronomy and Space Science Institute, Daejeon 34055, Republic of Korea}

\author[0000-0001-7594-8072]{Yongseok Lee}
\affil{Korea Astronomy and Space Science Institute, Daejeon 34055, Republic of Korea}
\affil{School of Space Research, Kyung Hee University, Yongin 17104, Republic of Korea}

\author{Dong-Joo Lee}
\affil{Korea Astronomy and Space Science Institute, Daejeon 34055, Republic of Korea}
\affil{Department of Astronomy and Space Science, Chungbuk National University, Cheongju 28644, Republic of Korea}

\author[0000-0002-8692-2588]{Kyeongsoo Hong}
\affil{Institute for Astrophysics, Chungbuk National University, Cheongju 28644, Republic of Korea}

\author[0000-0002-7511-2950]{Sang-Mok Cha}
\affil{Korea Astronomy and Space Science Institute, Daejeon 34055, Republic of Korea}
\affil{School of Space Research, Kyung Hee University, Yongin 17104, Republic of Korea}

\author{Dong-Jin Kim}
\affil{Korea Astronomy and Space Science Institute, Daejeon 34055, Republic of Korea}

\author{Byeong-Gon Park}
\affil{Korea Astronomy and Space Science Institute, Daejeon 34055, Republic of Korea}

\begin{abstract}
EL CVn-type eclipsing binaries are composed of a massive A-type main-sequence primary star and a hotter B-type secondary star. 
This paper presents the time-series photometric and asteroseismic results of the EL CVn-type star 1SWASP J024743.37-251549.2.
Well-defined eclipsing light curves were constructed by using the novel high-cadence $BV$ data and archival {\it TESS} data,
and the physical parameters of each binary component were derived by modeling the light curves.
Multiple frequency analysis was performed to investigate the pulsation properties of the binary components.
A reliable signal could not be detected in the high-frequency region of 100--300 day$^{-1}$, unlike in the previous discovery of three frequencies around 200 day$^{-1}$.
This indicates that the pulsation amplitudes of the pre-helium white dwarf secondary component decreased considerably.
By contrast, 12 frequencies were detected in the range of 33 to 53 day$^{-1}$. Most of them were classified as $\delta$ Sct-type pulsations originating from the primary star.
Theoretical frequencies for the seismic analysis were obtained by adding the non-rotating model frequencies from the GYRE and their rotational shifts from the complete calculation approach.
Grid-based fitting was conducted for various stellar properties. The theoretical frequencies and stellar parameters of the best solution concurred well with the observations.
The rotation rate was constrained to 1.50 $\pm$ 0.02 day$^{-1}$, indicating the synchronized rotation of the primary star.
The results imply that the complete approach based on the polytropic model is applicable to the seismic analysis of fast-rotating $\delta$ Sct stars.
\end{abstract}

\section{Introduction \label{sec_intro}}
EL CVn-type stars are a new type of eclipsing binaries introduced by \citet{maxted2014}. 
These binaries are composed of a massive A-type main-sequence primary star and a hotter B-type secondary star.
Their light curves show a boxy-shaped primary minimum and a slightly shallower secondary one, 
which are formed by the occultation and transit of a small secondary component, respectively.
The smaller, hotter star in the binary system implies that the star is highly evolved.
The massive progenitor of the present secondary star evolved earlier into the red giant phase and 
experienced a significant loss of mass through the non-conservative stable mass transfer driven by the Roche-lobe overflow \citep{chen2017}.
The stripped red giant star is now in a rarely observed state evolving to higher effective temperatures at a nearly constant luminosity,
before becoming an extremely low-mass white dwarf (ELM WD).
This ELM WD precursor, also known as a pre-He-WD, has a thick hydrogen envelope of about 0.005 $M_\sun$
surrounding the helium core \citep[hereafter referred to as Maxted13]{maxted2013},
where the helium could not burn because of an insufficient mass, lower than 0.3 $M_\sun$ \citep{heber2009, heber2016}.
However, the mass-gained primary component remained in the main-sequence phase.

The number of EL CVn-type binaries has increased considerably to approximately 70, owing to the photometric survey data.
For example, \citet{maxted2014} discovered 17 bright binaries with orbital periods of 0.7--2.2 days from the Wide Angle Search for Planets (WASP) database and 
\citet{vanRoestel2018} reported 36 new binaries with the periods of 0.46--3.8 days based on the Palomar Transient Factory (PTF) data. 
More than 15 such samples were discovered using the archival data of the {\it Kepler} or {\it TESS} space telescopes \citep{kwang2020}.
Our group has initiated the spectroscopic follow-up observations of the EL CVn stars to analyze their physical properties more accurately \citep{lee2020, hong2021}.

Some of these binaries were observed to exhibit multi-periodic pulsations in either one or both the components,
i.e., in the A-type primary stars of seven binaries and in the pre-He-WD secondary of five systems \citep{hong2021}.
Their stellar properties can be investigated comprehensively through asteroseismological analysis based on the pulsation frequencies \citep{brown1994, aerts2021}.

The eclipsing binary, 1SWASP J024743.37-251549.2 (hereafter referred to as J0247-25), was discovered by \citet{maxted2011} using the WASP photometric data.
They inferred that the small, hot companion is most likely a stripped red giant star whose mass was transferred to the A-type primary star, after comparing the observed parameters with various models.
Subsequently, Maxted13 performed ultra-high cadence photometric observations for three nights and detected multi-periodic pulsations for both the components.
\citet{istrate2017} conducted a seismic analysis of the pre-He-WD secondary based on the three frequencies detected by Maxted13
and obtained three possible solutions with different pulsation modes.

J0247-25 is the best-known binary discovered firstly among the EL CVn-type stars.
However, the pulsation properties of the binary components have not yet been sufficiently analyzed because of the limited photometric data.
In this study, we investigated the physical properties of J0247-25 using new extensive photometric data with a long observation period and high precision.

\section{Observation and Data Reduction \label{sec_obsdata}}
\subsection{$BV$ Photometry with KMTNet}
New $BV$ photometric data were obtained using the Korea Microlensing Telescope Network \citep[KMTNet,][]{kim2016a}.
The KMTNet system consists of three identical 1.6\,m telescopes installed at the Cerro Tololo Inter-American Observatory (CTIO) in Chile, 
the South African Astronomical Observatory (SAAO) in South Africa, and the Siding Spring Observatory (SSO) in Australia.
The observations of J0247-25 were conducted for 9 nights from July 28 to October 4, 2014, at the second host site, SAAO, 
and for 14 nights from November 1 to 22, 2014, at the third site, SSO, during the telescope commissioning phase of each site.
A 4k CCD camera with an image scale of 0.36 arcsec pixel$^{-1}$ and a field of view of 25$\times$25 arcmin$^2$ \citep{lee2016} was used.

The target field was monitored mainly with one filter $V$ to obtain high-cadence data, which are crucial for detecting the short-period variations of a highly dense object.
The exposure time was set to 10--15 seconds depending on the weather conditions and
by considering the pulsation periods of about 400 seconds for the pre-He-WD secondary star (Maxted13).
This resulted in the observing cadence of 17--22 seconds.

The time-series CCD images were processed with the IRAF package to correct the instrumental bias, dark noise, and pixel-to-pixel sensitivity (flat fielding).
Aperture photometry was applied after this pre-processing step to derive the instrumental magnitudes of the stars in the observed images.
The instrumental magnitudes of J0247-25 were standardized by using the photometric parameters of five stars around the variable target,
whose values were extracted from the AAVSO Photometric All Sky Survey \citep[APASS,][]{henden2016} catalog.
The transformation error is approximately 0.02 mag. Detailed explanations of the ensemble normalization method were presented in \citet{kim2016b}. 

\subsection{Archival Data from {\it TESS}}
J0247-25 was observed with the {\it TESS} in Sector 4 for 25.94 days from October 19 to November 14, 2018.
Its two-minute cadence data were downloaded from the Mikulski Archive for Space Telescopes (MAST, https://mast.stsci.edu).
The Pre-search Data Conditioning Simple Aperture Photometry (PDCSAP) flux, which has corrected long-term trends caused by instrumental effects, was adopted in this study.
The PDCSAP flux was converted into a magnitude in which the maximum brightness in the eclipsing light curve matches to the {\it TESS} magnitude $T$ =  12.111 \citep{stassun2019}.

Anomalous data with a non-zero "QUALITY" flag were removed.
Some bad data with a prominent discrepancy in the light curve ranging from BJD 2,458,421.216 to 2,458,421.401 and
with a significant scatter ranging from BJD 2,458,422.760 to  2,458,423.515 \citep[see median absolute deviation in figure 4 by][]{fausnaugh2019} were also excluded.

\section{Binary Properties \label{sec_binary}}
\subsection{Light Curve Analysis}
\begin{figure}
\center\includegraphics[scale=0.54]{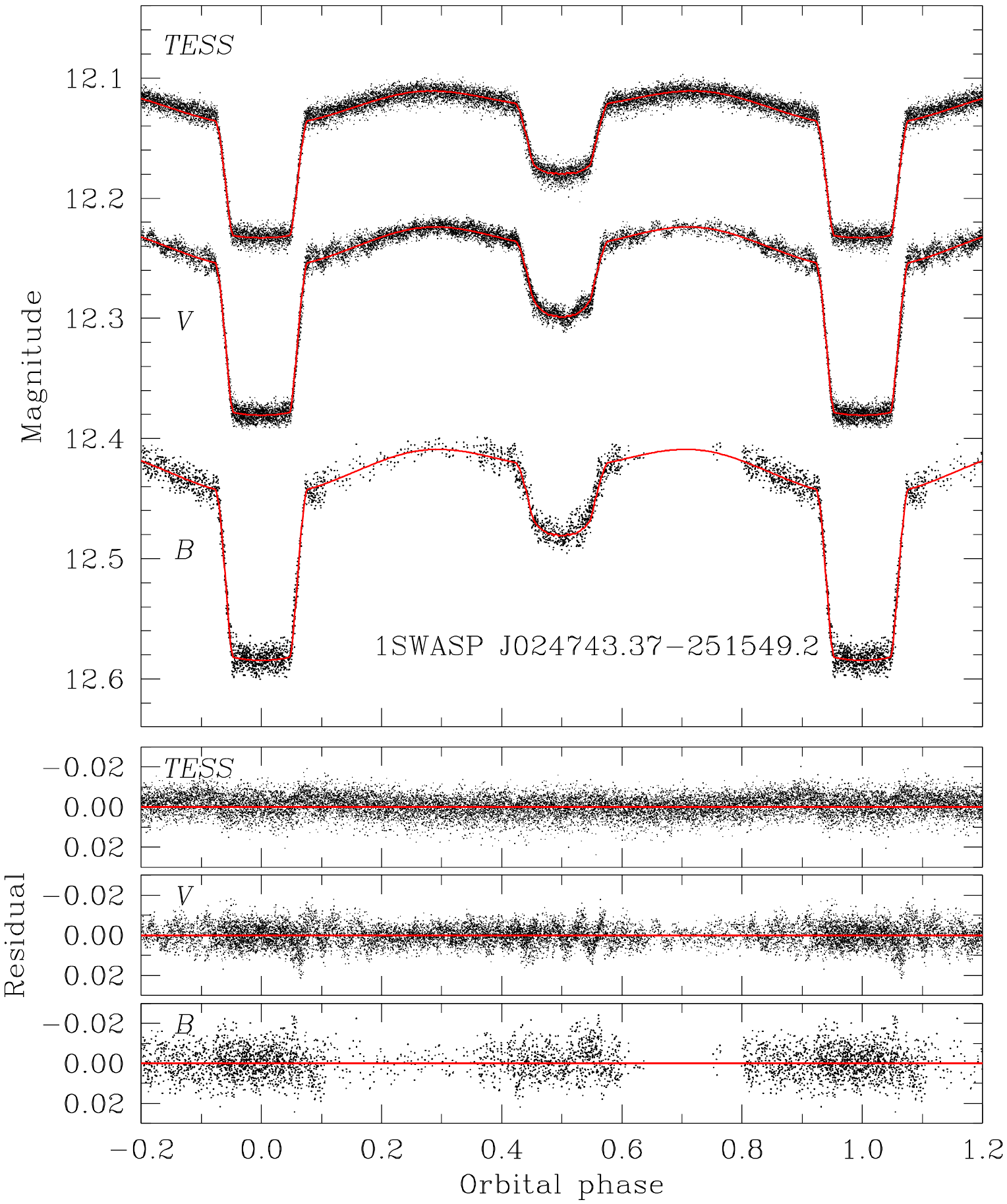}
\caption{Eclipsing light curves with fitted models (upper) and the residuals after subtracting the model curves from the observation data (lower).
The dots are individual measures from the {\it TESS} and KMTNet $BV$ observations. \label{fig_lc}}
\end{figure}

Figure \ref{fig_lc} presents the eclipsing light curves with the three data sets,
where the orbital phases were calculated with a period of 0.667830667 days and a minimum epoch of BJD 2,456,934.429445 (see Table \ref{tab_binary}).
The flat primary eclipse, which is a characteristic feature of the EL CVn-type binary systems \citep{maxted2014}, is well defined.
The curved shapes in the out-of-eclipse phases correspond to the ellipsoidal variations caused by the tidally elongated binary components.
The two maxima shown at the orbital phases $\sim$0.25 and $\sim$0.75, appear to be nearly symmetric.
This indicates that any surface inhomogeneity \citep[starspot,][]{hong2021} or Doppler boosting effect \citep{prsa2016} is negligible.
Additionally, the reflection effect due to the hotter star, i.e., brightening around the secondary eclipse, was observed with greater visibility in the shorter wavelength $B$ band.

All three datasets were simultaneously modeled using the 2007 version of the Wilson--Devinney synthesis code \citep[hereafter W--D]{wilson1971, vanHamme2007}. 
The mass ratio was set to $q$ = $M_{\rm B}$/$M_{\rm A}$ = 0.1371 $\pm$ 0.0012
from the velocity semi-amplitudes of $K_{\rm A}$ = 33.9 $\pm$ 0.3 km sec$^{-1}$ and $K_{\rm B}$ = 247.2 $\pm$ 0.4 km sec$^{-1}$ by Maxted13. 
The subscript A represents the primary star, J0247-25A, and B represents the secondary star, J0247-25B.
The initial values of the effective temperature were obtained from Maxted13. 
Other parameters such as bolometric albedos, gravity-darkening exponents, and limb-darkening coefficients were obtained using the same method as in \citet{lee2012}.

\begin{deluxetable}{lcc}
\tablewidth{0pt}
\tablecaption{Binary parameters of J0247-25 \label{tab_binary}}
\tablehead{ \colhead{Parameter}  & \colhead{Primary} & \colhead{Secondary} }
\startdata                                                                         
$T_0$ (BJD)                              & \multicolumn{2}{c}{2,456,934.429445$\pm$0.000049}       \\
$P$ (day)                                & \multicolumn{2}{c}{0.667830667$\pm$0.000000046}         \\
$q$                                        & \multicolumn{2}{c}{0.1371$\pm$0.0012}                   \\
$i$ (deg)                                & \multicolumn{2}{c}{85.87$\pm$0.13}                      \\
$T_{\rm eff}$ (K)                        & 7,760$\pm$200               & 10,590$\pm$400               \\
$\Omega$                                 & 2.819$\pm$0.010           & 3.049$\pm$0.010             \\
$L/(L_1+L_2)_{B}$                        & 0.8878$\pm$0.0024         & 0.1122                      \\
$L/(L_1+L_2)_{V}$                        & 0.8988$\pm$0.0026         & 0.1012                      \\
$L/(L_1+L_2)_{T}$                        & 0.9210$\pm$0.0023         & 0.0790                      \\
$r_{\tt pole}$                               & 0.3717$\pm$0.0011         & 0.0846$\pm$0.0006           \\
$r_{\tt point}$                              & 0.3905$\pm$0.0014         & 0.0854$\pm$0.0006           \\
$r_{\tt side}$                               & 0.3835$\pm$0.0013         & 0.0848$\pm$0.0006           \\
$r_{\tt back}$                              & 0.3873$\pm$0.0013         & 0.0853$\pm$0.0006           \\
$r_{\tt volume}$$^\dagger$                     & 0.3809$\pm$0.0013         & 0.0849$\pm$0.0006           \\ [1.0mm]
\multicolumn{3}{l}{Absolute parameters:}                                                           \\
$a$ ($R_\odot$)                          & \multicolumn{2}{c}{3.719$\pm$0.007}                     \\            
$M$ ($M_\odot$)                          & 1.362$\pm$0.007          & 0.187$\pm$0.002              \\
$R$ ($R_\odot$)                          & 1.416$\pm$0.005          & 0.316$\pm$0.002              \\
$\log$ $g$ (cgs)                         & 4.270$\pm$0.003          & 4.711$\pm$0.007              \\
$L$ ($L_\odot$)                          & 6.5$\pm$0.7              & 1.1$\pm$0.2                  \\
$M_{\rm bol}$ (mag)                      & 2.7$\pm$0.1              & 4.6$\pm$0.2                  \\
\enddata
\tablenotetext{^\dagger}{Mean volume radius.} 
\end{deluxetable}

This synthesis was repeated until the corrections of the adjustable parameters became smaller than their standard deviations,
using the differential correction program of the W--D code.
The orbital eccentricity was set as a free parameter, but the value retained zero, implying a circular orbit.
The final results are listed in Table \ref{tab_binary} and the synthetic light curves are plotted as the solid curves in Figure \ref{fig_lc}.  
Following the procedure applied by \citet{koo2014}, the errors for the adjustable parameters were obtained 
by splitting the observed data into five subsets and analyzing them individually with the W--D code. 

\subsection{Physical Parameters \label{subsec_phy_par}}
The absolute dimensions for both components were derived using the JKTABSDIM code \citep{southworth2005} 
based on our photometric parameters and the previous spectroscopic results ($K_{\rm A}$ and $K_{\rm B}$ by Maxted13). 
These are presented in the lower part of Table \ref{tab_binary}.
The luminosity and bolometric magnitude were computed by adopting $T_{\rm eff,\sun}$ = 5,776 K and $M_{\rm bol,\sun}$ = +4.73 for the solar values.
It was assumed that the temperatures of the primary and secondary components have errors of 200 K and 400 K, respectively,
as given by Maxted13 because the temperature errors yielded by the W--D code are underestimated.

The radius and related parameters such as luminosity and gravity, which are presented in Table \ref{tab_binary}, vary significantly from the previous results obtained by Maxted13.
The radii of the primary and secondary components, 1.416 $\pm$ 0.005 $R_\sun$ and 0.316 $\pm$ 0.002 $R_\sun$, are 17\% and 14\% smaller than
the previous values of 1.697 $\pm$ 0.011 $R_\sun$ and 0.368 $\pm$ 0.005 $R_\sun$, respectively.

A distance of 847 $\pm$ 47 pc was estimated by using the absolute bolometric magnitude $M_{\rm bol}$ = 2.7 $\pm$ 0.1
and the apparent magnitude $V$ = 12.38 $\pm$ 0.02 for the primary star.
The apparent magnitude was derived from the data during the total-eclipsed phase when only the primary component was visible.
The interstellar reddening was adopted as $E(B-V) = 0.01$ (Maxted13), and the bolometric correction $BC = -0.01$ was deduced from \citet{popper1980}.

The estimated distance value of 847 $\pm$ 47 pc is notably lower than the previous value of 1,035 $\pm$ 55 pc given by Maxted13,
but concurs well with the recent results based on Gaia Early Data Release 3 \cite[Gaia Collaboration,][]{brown2021}.
The reciprocal of Gaia's parallax 1.1977 $\pm$ 0.0182 mas is 835 $\pm$ 13 pc and 
the photogeometric distance estimated by \citet{bailer2021} is 807$^{+12}_{-10}$ pc.

\begin{figure}
\center\includegraphics[scale=0.61]{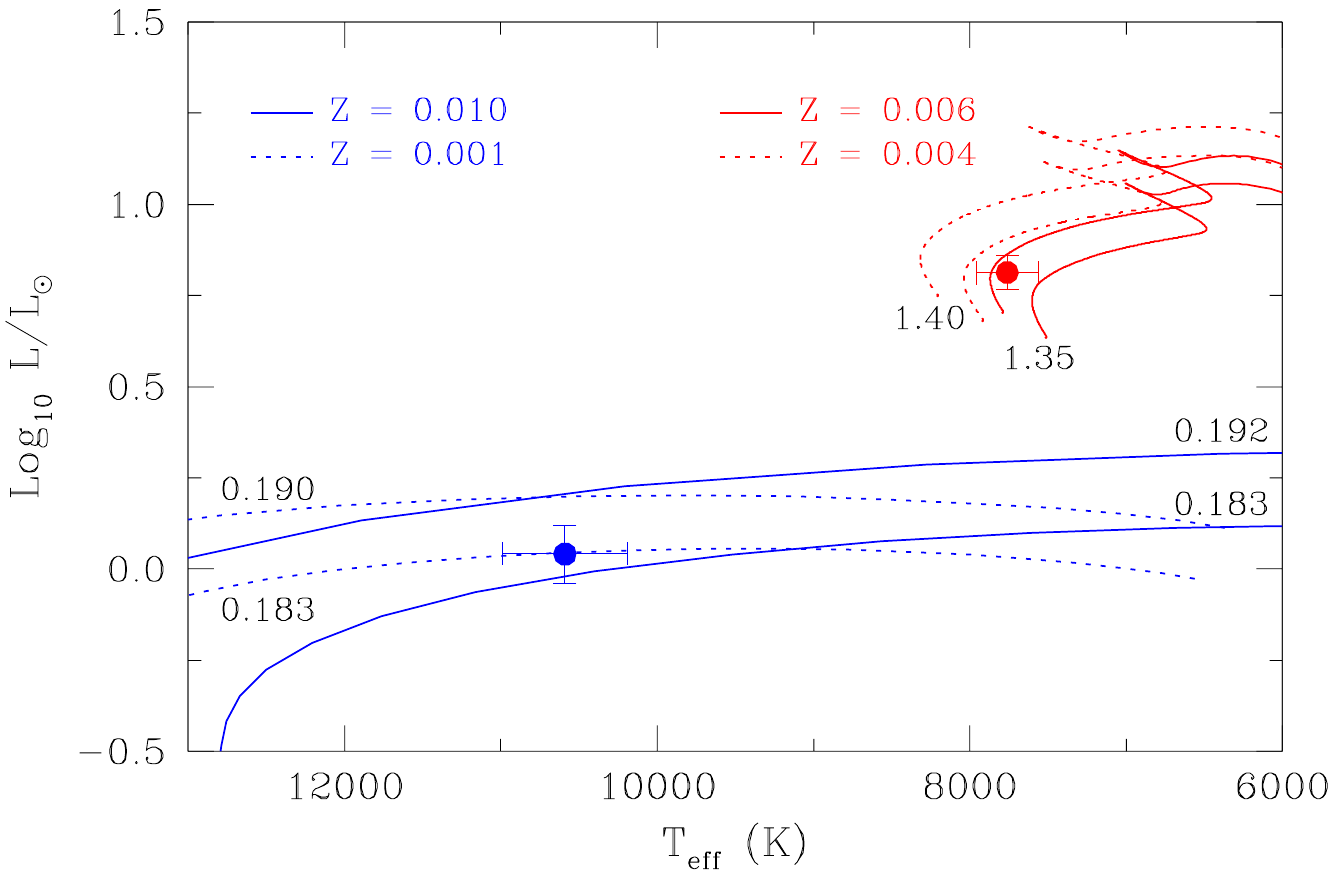}
\caption{Position of J0247-25 in the HR diagram, along with theoretical evolutionary tracks for given stellar masses in solar unit and metallicities in Z.
The red and blue dots with error bars represent the primary and secondary stars, respectively.
Red lines are stellar models of the main-sequence phase by \citet{bressan2012}, and the blue lines are models of ELM WDs by \citet{istrate2016}. \label{fig_hr}}
\end{figure}

Figure \ref{fig_hr} shows that the parameters determined in this study are in good agreement with those of the theoretical evolution models in the Hertzsprung-Russell (HR) diagram. 
The metallicity appears to be Z $\approx$ 0.005, which concurs well with the kinematic estimation of [Fe/H] = $-0.65 \pm 0.35$ by \citet{maxted2011}.
Additionally, the gravity of J0247-25B, $\log g$ = 4.711 $\pm$ 0.007 cm sec$^{-2}$, is larger than the previous value of 4.576 $\pm$ 0.011 cm sec$^{-2}$,
but is very close to the spectroscopic estimation of 4.70$^{+0.11}_{-0.12}$ cm sec$^{-2}$ \citep{istrate2017}.

Consequently, it is presumed that the physical parameters obtained in this study are more reliable than those obtained by Maxted13.
In contrast to our extensive data, they used limited data for only three nights. The short-term light variations by pulsations
appeared to interfere with the determination of the accurate times of eclipse ingress and egress, directly affecting their radius estimation.

\section{Pulsation Properties \label{sec_pulsation}}
\begin{figure}
\center\includegraphics[scale=0.56]{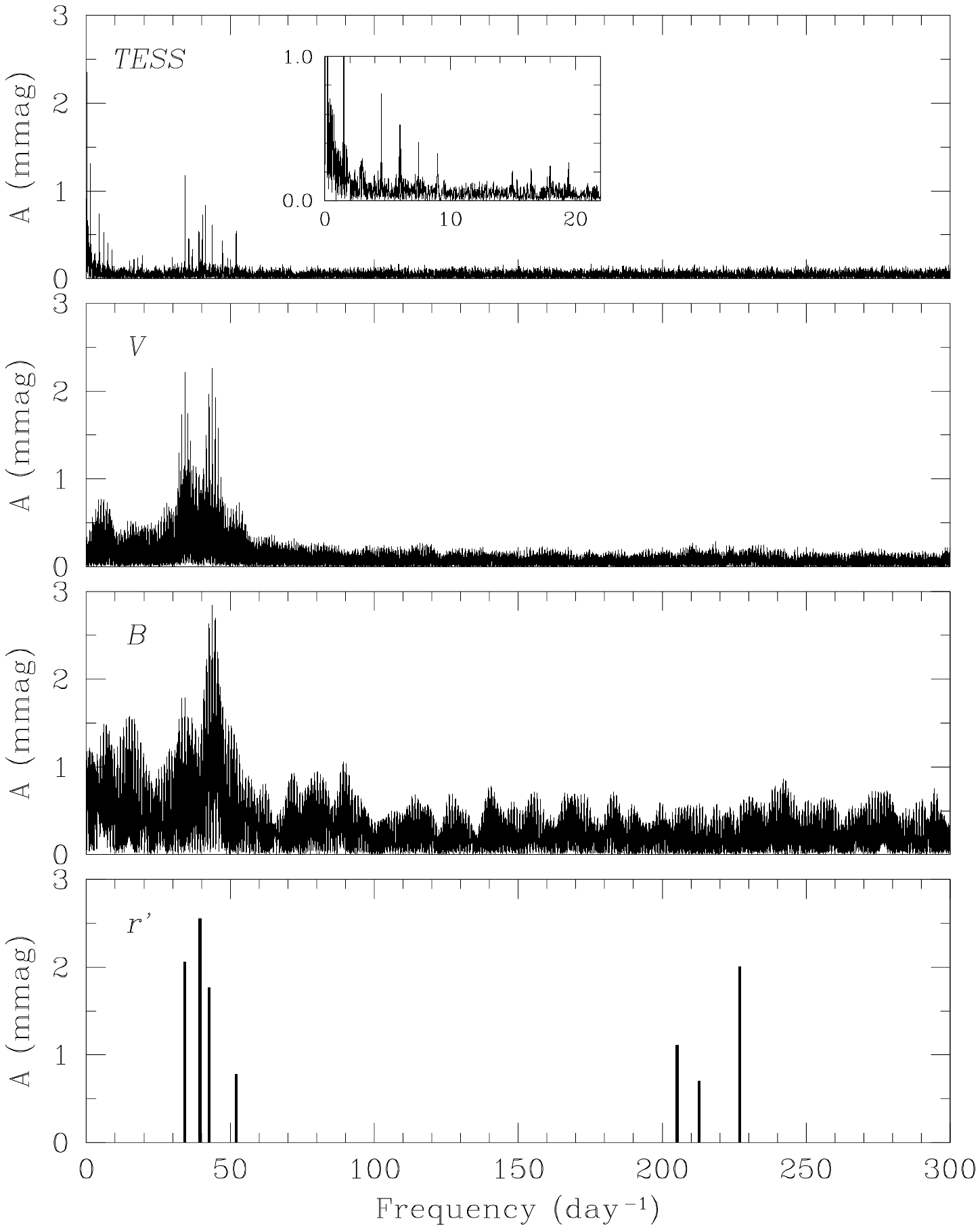}
\caption{Fourier amplitude spectra with the {\it TESS}, KMTNet $V$, and $B$ band data, in order from the top.
A schematic diagram of the previous results by Maxted13 is also shown at the bottom. The same amplitude scale of 0.0--3.0 mmag is set for comparison.
The insert at the top is a closer perspective view which shows a series of orbital frequency harmonics in the frequency region less than 22 day$^{-1}$. \label{fig_full_spectra}}
\end{figure}

A multiple frequency analysis was conducted, which applies the discrete Fourier transform and the least-squares fitting process \citep{kim2010}.
Figure \ref{fig_full_spectra} shows the Fourier amplitude spectra of the residuals after subtracting the model light curves from the observation data.
As expected, the space-based 24-hour coverage {\it TESS} data showed the lowest amplitude noise and clear peaks without any sidelobe.
Conversely, the scarce KMTNet $B$ band data generated noise-rich spectra and severe aliasing sidelobes, which were inadequate for a detailed analysis.

Several peaks in the low-frequency region of the {\it TESS} amplitude spectra were found to be orbital frequency and its harmonics, 
i.e.,  $f_{orb}$ = 1.497 day$^{-1}$, $2f_{orb}$, $3f_{orb}$,  $4f_{orb}$,  $5f_{orb}$, $6f_{orb}$, $10f_{orb}$, $11f_{orb}$, $12f_{orb}$, and also $13f_{orb}$ = 19.466 day$^{-1}$. 
The $Nf_{orb}$ harmonics can be explained by tidally excited oscillations (TEOs) or the imperfect removal of the eclipsing light curve \citep{guo2019}.
The TEOs are excited in a close binary system with high eccentricity \citep[KOI-54 as a representative example,][]{welsh2011},
but both the radial velocity data (Maxted13) and our binary analysis favored a circular orbit of J0247-25.
Therefore, the TEOs were not considered further, and light variations of all the orbital harmonics up to $47f_{orb}$ = 70.377 day$^{-1}$ were removed 
before the pre-whitening process presented in Section \ref{subsec_pul_pri}.

Several frequencies less than $f_{orb}$ were also detected. These low frequencies are either instrumental artifacts or originate from the binary star.
The $\gamma$ Dor-type gravity-mode pulsations of the A-type primary star are a strong candidate for stellar origin.
However, the physical properties listed in Table \ref{tab_binary} indicate that the primary component is hotter than the blue edge of the $\gamma$ Dor instability strip \citep{dupret2005};
thus, it may be difficult to anticipate the $\gamma$ Dor-type low frequencies.
Accordingly, the detailed analysis presented in the following subsections focuses on the high- and intermediate-frequency regions.

\subsection{Pulsations in the ELM pre-He-WD Component}
The most striking feature in Figure \ref{fig_full_spectra} is the absence of amplitude peaks in the high-frequency region near 200 day$^{-1}$ for all three datasets.
As shown at the bottom of the figure, Maxted13 discovered three frequencies, i.e., 226.8, 205.4, and 212.9 day$^{-1}$
with the observed amplitudes of 2.0, 1.1, and 0.7 mmag in the $r'$ band, respectively.
These amplitudes are sufficiently large to enable detection, considering the noise levels of our data.

\begin{figure}
\center\includegraphics[scale=0.56]{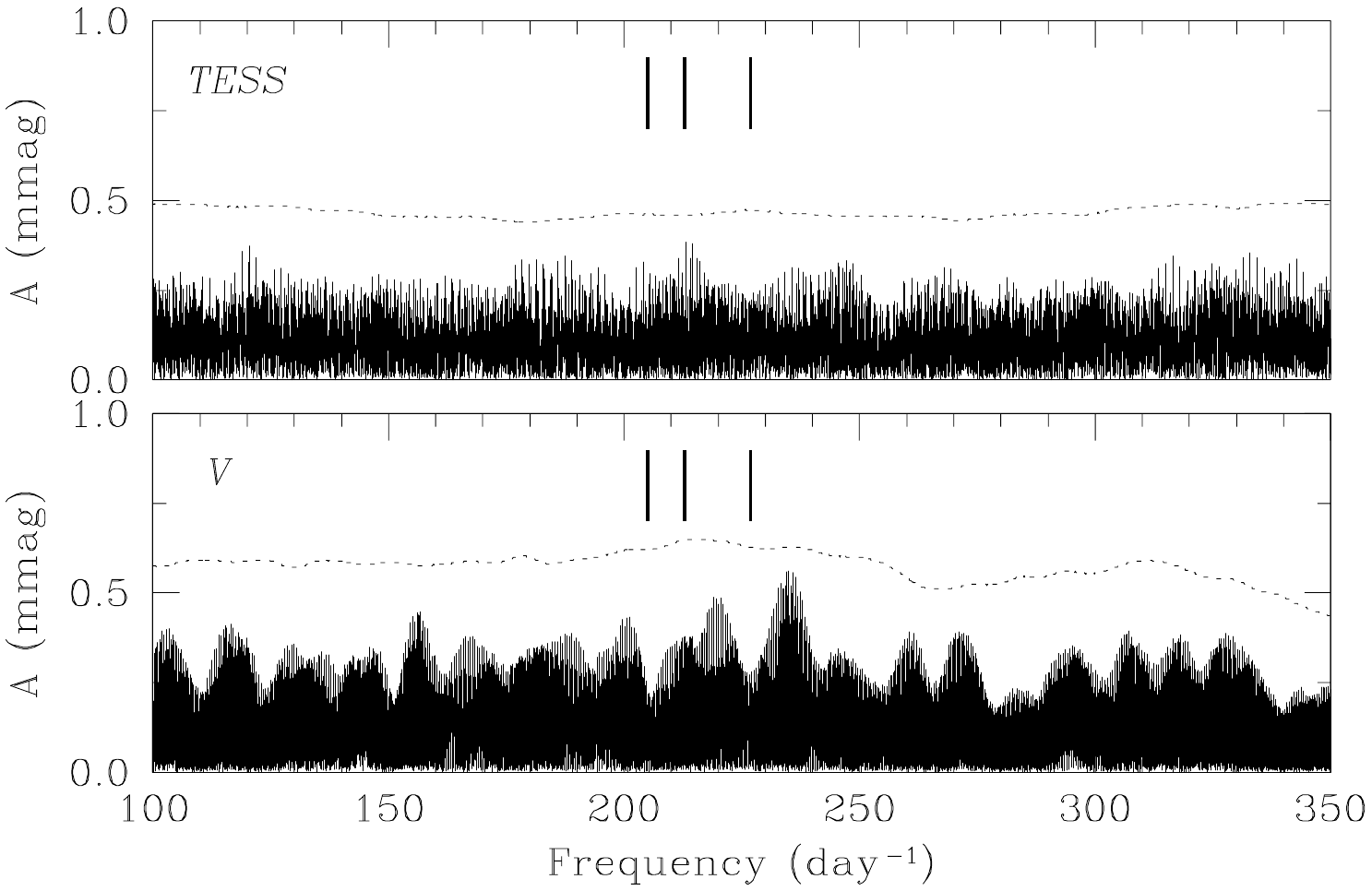}
\caption{Fourier spectra in the high-frequency region using only the data around the secondary eclipse phase.
The three vertical lines represent the pulsation frequencies detected previously by Maxted13, with the observed amplitudes higher than 0.7 mmag in the $r'$ band.
The dotted lines show the signal level, applying four times the noise,
which is calculated by averaging the amplitudes for $\pm$25 day$^{-1}$ boxes around each frequency. \label{fig_high_spectra}}
\end{figure}

Maxted13 identified these frequencies as pulsations excited in the pre-He-WD secondary component, J0247-25B.
However, our amplitude spectra, shown in Figure \ref{fig_full_spectra}, were obtained with the data including the primary eclipse
when the pre-He-WD was invisible or obscured by the A-type primary component.
Therefore, following the analysis of Maxted13, the amplitude spectra were recalculated using only the data around the secondary eclipse phase, 
where the secondary star has the highest flux contribution among the entire orbital phase.
Figure \ref{fig_high_spectra} shows the new amplitude spectra in the high-frequency region of 100--350 day$^{-1}$; the Nyquist frequency of the {\it TESS} data is 360 day$^{-1}$. 
The noise levels were slightly increased because of the reduction in the analyzed data. 

A reliable peak with an amplitude larger than about 0.6 mmag, which is four times the noise level, could not be identified even in the new spectra.
Two possible explanations can be presented from an observational perspective.
Firstly, pulsations in the pre-He-WD showed wavelength-dependent amplitudes (Maxted13), being the largest in the shortest wavelength $u'$ band.
The central wavelength of the $r'$ band, $\lambda_{cen} =  6,260$ {\AA}, is shorter than that of the {\it TESS} filter ($\lambda_{cen} \simeq$ 8,000 {\AA}) 
but longer than that of the KMTNet $V$ band ($\lambda_{cen} \simeq$ 5,500 {\AA}).
Hence, the wavelength dependency does not account for the absence of amplitude peaks.

Secondly, the {\it TESS} data were obtained with a cadence of two minutes.
The cadence may not be sufficiently high to analyze rapid light variations with a short periodicity of seven minutes ($\approx$ 200 day$^{-1}$) for J0247-25B.
The low cadence causes amplitude suppression induced by flux integration.
The suppression rate can be estimated from the expression, sinc($\pi \nu / \nu_{samp}$), where $\nu$ is the pulsation frequency and $\nu_{samp}$ is the sampling rate
\citep{antoci2019}; if $\nu_{samp}$ = $\nu$, the value is zero, indicating that the amplitude is perfectly suppressed (or no light variation).
The suppression rate for the {\it TESS} data is 87\%, which is insufficient for the absence of amplitude peaks.
Furthermore, the KMTNet $V$ band data have a high cadence of $\approx$20 seconds, producing almost no suppression.
Therefore, it can be inferred that the observation cadence does not account for the lack of amplitude peaks.

Consequently, our results indicate that the pulsation amplitudes of the pre-He-WD secondary component decreased significantly.
\citet{istrate2017} also noted that the pulsations could not be detected on one night in 2014 when the follow-up observations of J0247-25 were made with the SOAR telescope.

\subsection{Pulsations in the A-type Main-Sequence Star \label{subsec_pul_pri}}
In Figure \ref{fig_full_spectra}, the dominant peaks for our three datasets are observed in the intermediate-frequency region of 30--55 day$^{-1}$.
As shown at the bottom of the figure, four frequencies were also found in this region by Maxted13,
who identified them as $\delta$ Sct-type pulsations excited in the A-type main-sequence primary star.

A $\delta$ Sct star has different pulsation amplitudes based on the wavelength (or passband),
maximizing at around 4,000 {\AA} and decreasing down to nearly half at around 7,000 {\AA} \citep{balona1999}.
This may be the reason why the pulsation amplitudes of the {\it TESS} data are overall smaller (approximately half) than those of the KMTNet $V$ and $B$ band data.

The dominant frequency with the highest amplitude was different between the datasets, i.e., 34.2 day$^{-1}$ for the {\it TESS}, 
43.7 day$^{-1}$ for the KMTNet $V$ and $B$, and 39.4 day$^{-1}$ for the previous Maxted13 data.
The amplitude variation occurred in most of the pulsation frequencies (see Table \ref{tab_freq}), and this is the observational characteristic of $\delta$ Sct stars \citep{breger2000}.

\begin{figure}
\center\includegraphics[scale=0.56]{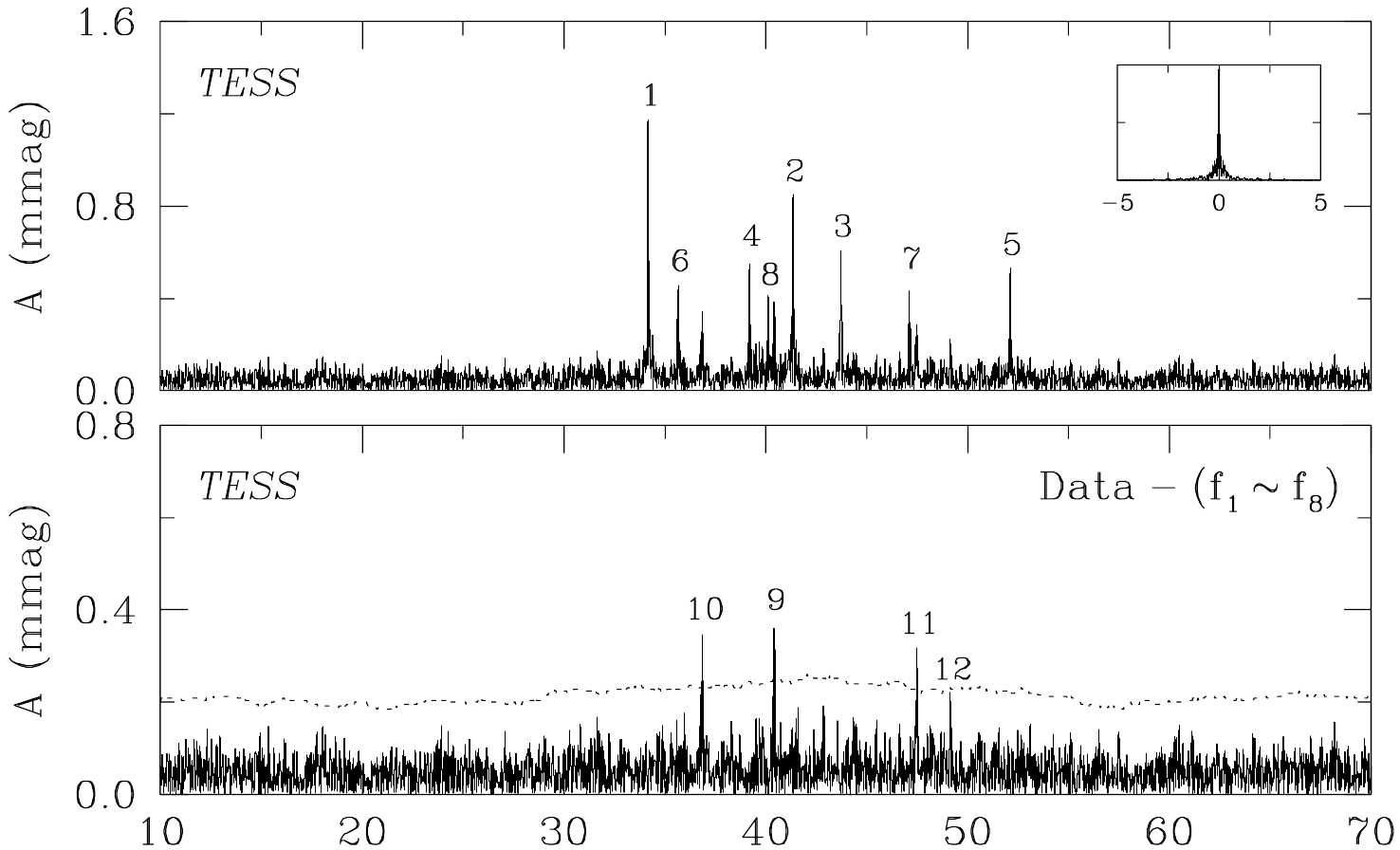}
\includegraphics[scale=0.56]{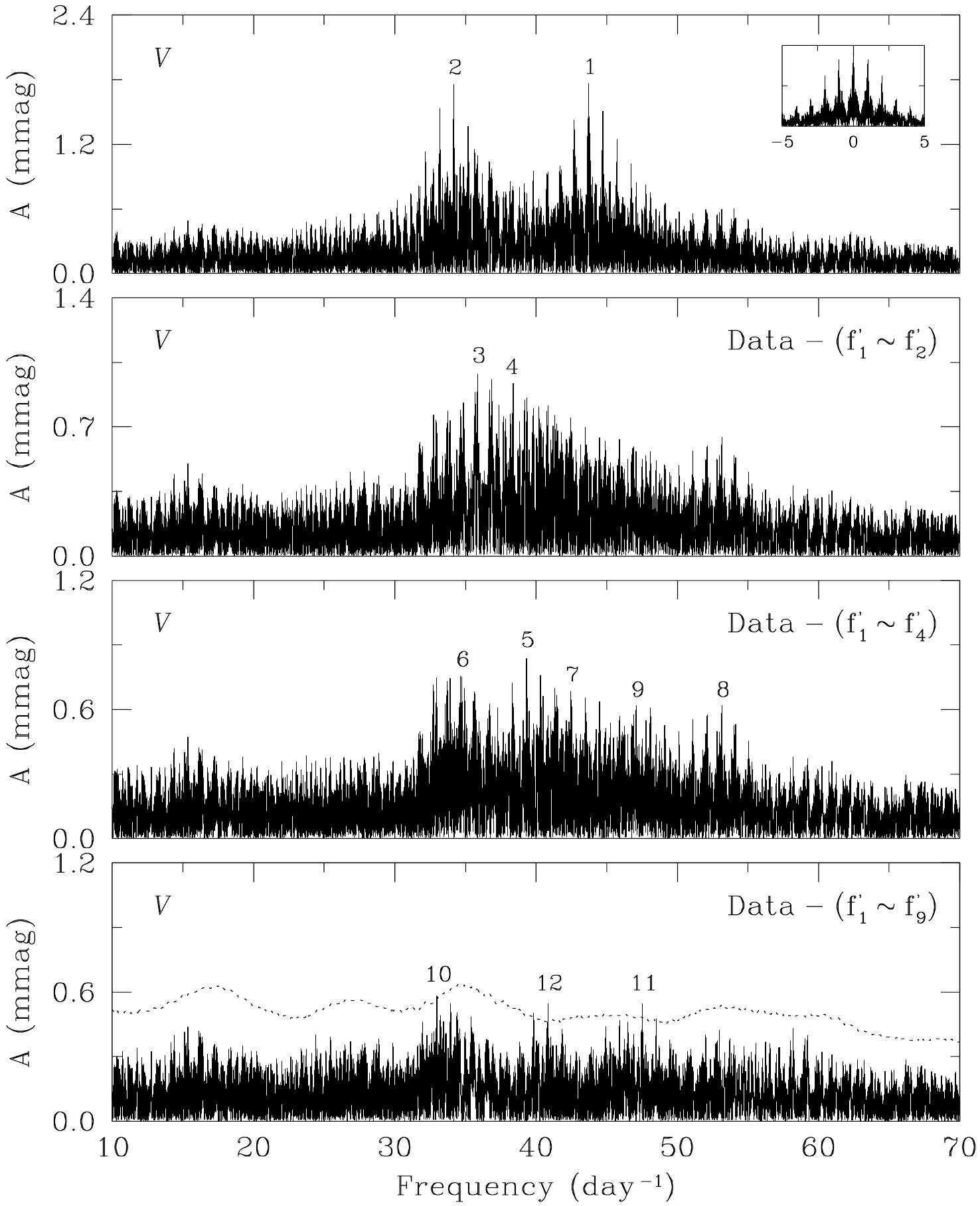}
\caption{Fourier spectra in the intermediate-frequency region. The upper two spectra were derived from the {\it TESS} data and the lower four with the KMTNet $V$ band data.
The inset panels show the window spectra for each dataset. Numbers (i) are the sequence of frequencies detected from the consecutive pre-whitening process.
The dotted lines represent four times the noise calculated with $\pm$2.5 day$^{-1}$ boxes around each frequency after subtracting the final frequency. \label{fig_med_spectra}}
\end{figure}

\begin{deluxetable*}{rrcrrl}
\tablewidth{0pt} \tablecolumns{6}
\tablecaption{Multiple frequencies detected from the A-type primary star \label{tab_freq}}
\tablehead{ 
\multicolumn{2}{c}{{\it TESS} (data in 2018)} & & \multicolumn{3}{c}{KMTNet $V$ band (data in 2014)}  \\
\cline{1-2} \cline{4-6} 
\colhead{Frequency$^\dagger$} & \colhead{Amplitude$^\dagger$} & &
\colhead{Frequency$^\dagger$} & \colhead{Amplitude$^\dagger$} & \colhead{Remark}  } 
\startdata
    $f_1$ = 34.176(1) & 1.17(5) &   & $f'_1$ = 43.7135(1) & 1.98(5)  & $= f_3$ \\
    $f_2$ = 41.346(1) & 0.84(5) &   &  $f'_2$ = 34.1759(1) & 1.83(6)  & $= f_1$  \\
    $f_3$ = 43.713(2) & 0.60(5) &   &  $f'_3$ = 35.8615(2) & 1.12(5)  & $= f_{10} - 1.0$ \\
    $f_4$ = 39.195(2) & 0.56(5) &   &  $f'_4$ = 38.1990(3) & 0.94(5)  & $= f_4 - 1.0$ \\
    $f_5$ = 52.120(2) & 0.52(5) &   &  $f'_5$ = 39.3251(2) & 0.97(5)  & $\simeq f_2 - 2.0$ \\ 
    $f_6$ = 35.669(2) & 0.45(5) &   &  $f'_6$ = 34.7303(2) & 0.79(5)  & $\approx f_6 - 1.0$ \\       
    $f_7$ = 47.117(2) & 0.46(5) &   &  $f'_7$ = 42.4706(2) & 0.78(5)  & $\approx f_9 + 2.0$\\
    $f_8$ = 40.125(2) & 0.39(5) &  &  $f'_8$ = 53.1371(3) & 0.64(5)  & $\simeq f_5 + 1.0$ \\ 
    $f_9$ = 40.402(2) & 0.37(5) &  &  $f'_9$ = 47.1050(3) & 0.65(5)  & $\simeq f_7$          \\ 
$f_{10}$ = 36.862(3) & 0.35(5) &  &  $f'_{10}$ = 32.9626(3) & 0.67(6)  & $\approx f'_3 - 2f_{orb}$ \\
$f_{11}$ = 47.481(3) & 0.32(5) &  &  $f'_{11}$ = 47.5098(4) & 0.55(5)  & $\simeq f_{11}$      \\  
$f_{12}$ = 49.127(5) & 0.22(5) &  &  $f'_{12}$ = 40.8386(3) & 0.57(5)  & $\simeq f'_5 + f_{orb}$ \\  
\enddata
\tablenotetext{^\dagger}{The units of the frequency and amplitude are day$^{-1}$ and mmag, respectively. The values in parentheses are errors in the last digit.}
\end{deluxetable*}

Multiple frequencies of the $\delta$ Sct-type star were obtained from the consecutive pre-whitening process using the eclipse-subtracted residuals for all the orbital phases.
The dilution or obscuration effect due to the secondary star was assumed to be negligible; if any, it produces false $f_{orb}$ splitting for a given pulsation frequency \citep{kim2010}.
Figure \ref{fig_med_spectra} presents the amplitude spectra of the {\it TESS} and KMTNet $V$ band data in the intermediate-frequency region of 10--70 day$^{-1}$.
The {\it TESS} and KMTNet frequencies are represented as $f_i$ and $f'_i$, respectively.
The window spectra of the KMTNet data showed conspicuous sidelobes of the 1.0 day$^{-1}$ aliasing.

We detected 12 frequencies from each dataset by applying the empirical criterion of the signal-to-noise amplitude ratio S/N $\ge$ 4.0 \citep{breger1993}.
As listed in Table \ref{tab_freq}, several KMTNet frequencies exhibit a difference of 1.0 or 2.0 day$^{-1}$ from the {\it TESS} ones,
that is, $f'_3$, $f'_4$, $f'_5$, $f'_6$, $f'_7$, and $f'_8$, which are due to the 1.0 day$^{-1}$ aliasing effect.
The {\it TESS} frequencies were considered as accurate values without contamination from aliasing.

For the 12 {\it TESS} frequencies, the linear combination terms were analyzed with the orbital frequency, $f_i = f_j \pm m f_{orb}$, or with the other frequencies, $f_i = m f_j \pm n f_k$.
Here, the numbers, $m$ and $n$,  are integers.
Four frequencies were found to be orbital harmonic combinations: $f_6 = f_1 + f_{orb}$, $f_8 = f_5 - 8 f_{orb}$, $f_{10} = f_2 - 3f_{orb}$,  and $f_{12} = f_5 - 2 f_{orb}$.
The differences between the observed and predicted values are 0.004, 0.016, 0.008, and 0.002 day$^{-1}$, respectively,
which are significantly smaller than the Rayleigh resolution criterion of 1/$\Delta$T = 1/25.94 = 0.039 day$^{-1}$.
Among these, $f_{10}$ was reclassified as a pulsation frequency excited in the primary star
because the KMTNet $V$-band amplitude of the corresponding frequency $f'_3$ ($= f_{10} - 1.0$) is too large.

Additionally, the frequency, $f_9$ = 40.402 day$^{-1}$, is close to the orbital harmonic, $27f_{orb}$ = 40.429 day$^{-1}$,
the difference of which is smaller than the Rayleigh criterion.
Given that all the orbital harmonics, including $27f_{orb}$ with an amplitude of 0.59 mmag, have already been subtracted,
$f_9$ with an amplitude of 0.37 mmag should be a pulsation frequency separated from $27f_{orb}$.
Before we removed the harmonics, these two frequencies were detected as a single frequency of 40.418 day$^{-1}$ with an amplitude of 0.72 mmag.

\section{Seismic Analysis and Rotation Effect \label{sec_modeling}}
The observed frequencies of the primary star were compared with the theoretical ones.
Nine frequencies, except for the three combination frequencies of $f_6$, $f_8$, and $f_{12}$, were used initially.
During the seismic analysis, $f_6$ matched excellently with the theoretical frequency.
$f_6$ was then considered as a pulsation frequency, and the analysis was performed again with 10 frequencies, excluding $f_8$ and $f_{12}$.
The best solutions with and without $f_6$ are identical, but the former shows a better fit, that is, a lower $\chi^2$. 

The evolution of a single star was applied, considering that the theoretical frequencies of a binary component are very close to those of a single star
with the same physical properties of mass, radius, and metallicity \citep{streamer2018}.
Age was not considered in this paper because it is different between the binary component and a single star owing to binary interactions such as mass transfer.

\subsection{Evolutionary and Pulsation Models}
Stellar evolutionary models were obtained using the Modules for Experiments in Stellar Astrophysics \citep[MESA,][]{paxton2011, paxton2013, paxton2015, paxton2018, paxton2019} software version 11701.
The initial metal fraction was set to use the recent solar composition given in AGSS09 \citep{asplund2009}, {\tt initial\_zfracs = 6},
and the opacity table was set to the corresponding composition, {\tt kappa\_file\_prefix = 'a09'} and {\tt kappa\_lowT\_prefix = 'lowT\_fa05\_a09p'}.
The modified mixing length theory (MLT) of \citet{henyey1965} was adopted,
and the mixing length parameter was set to $\alpha_{\rm MLT}$ = 1.0, deduced from the metallicity-dependent relation by \citet{viani2018}.
Most of the other parameters were used in the default settings of the MESA.

The rotation effect was not considered in this stellar modeling.
\citet{georgy2013} found that the evolution track of 1.0 $M_\sun$ remains nearly unchanged by rotation
and that the rotating model reaches a higher luminosity at the end of the main-sequence phase for a higher mass.
The evolution tracks with and without rotation look very similar to each other near the zero-age main-sequence phase.

The theoretical pulsation frequencies were obtained for these stellar models without rotation using the GYRE code \citep{townsend2013}, which is distributed with the MESA.
The adiabatic calculation was adopted, and the modes with radial orders $n$ = 1--10 and angular degrees $\ell$ = 0--3 were considered.
The stellar model parameters were set as {\tt model\_type = 'EVOL'} and {\tt file\_format = 'MESA'} to read the MESA's GYRE-format output file.
The fourth-order Gauss-Legendre collocation scheme was used for solving differential equations numerically, {\tt diff\_scheme = 'COLLOC\_GL4'}.
The grid parameters were set as {\tt alpha\_osc = 10} and {\tt alpha\_exp = 2} \citep{murphy2021}.
Most of the remaining parameters were left to their default values.

\subsection{Rotational Shift of Pulsation Frequencies \label{sec_rotshift}}
The pulsation frequency shifts by stellar rotation due to both the Coriolis and centrifugal forces \citep{saio1981, perezhernandez1995, soufi1998, goupil2000}.
However, the GYRE code incorporates only an approximate treatment of the Coriolis force
because solving the pulsation equations for a rotating star is very complicated\footnote{Documentation for the GYRE, https://gyre.readthedocs.io/}.
Therefore, we obtained the pulsation frequencies for non-rotating models and then added the frequency shifts estimated from the other numerical approach.

\begin{figure}
\center\includegraphics[scale=0.51]{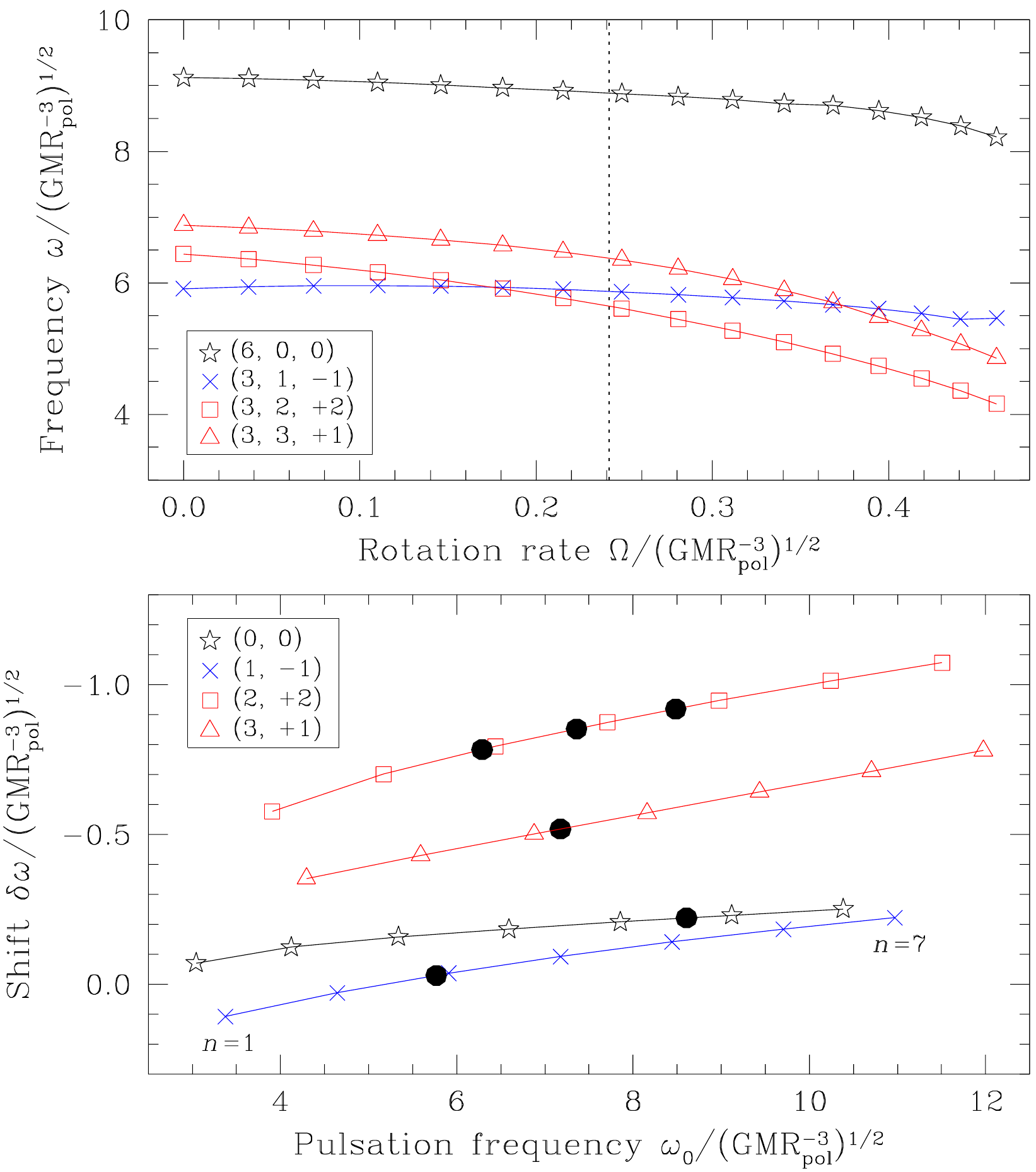}
\caption{(Upper) Variations of the pulsation angular frequency with increasing rotation rate.
These data were obtained from the complete calculation by \citet{reese2006a}. The symbols represent the pulsation modes ($n$, $\ell$, $m$).
The vertical dashed line represents the rotation rate $\Omega = 0.2414$, corresponding to the best solution described in Section \ref{sec_solution}. 
(Lower) The frequency shift $\delta\omega$ versus $\omega_0$ with the radial order $n$ = 1--7. 
The six black dots represent the pulsation frequencies from the GYRE, and their frequency shifts were obtained by interpolating the polytropic values of two adjacent orders.
The other symbols denote the modes ($\ell$, $m$). \label{fig_rot_shift}}
\end{figure}

If the rotational shift of the pulsation frequency is defined as $\delta\omega \equiv \omega - \omega_0$, 
where $\omega$ is the angular frequency observed in an inertial frame and $\omega_0$ is the frequency corresponding to the non-rotating case,
the value of $\delta\omega$ varies based on the pulsation mode ($n$, $\ell$, $m$)\footnote{The stellar rotation causes the non-radial ($\ell \ne 0$) modes to split into
a series of azimuthal orders, $m$, from $-\ell$ to $+\ell$. This article follows the convention that positive $m$ is the retrograde mode and negative $m$ is the prograde mode.}
as well as the rotation rate $\Omega$ (see Figure \ref{fig_rot_shift}).
The rotational shift was analyzed up to the third order, $\Omega^3$, through the perturbative approach \citep{saio1981, goupil2000}.

We applied the non-perturbative complete calculation introduced by \citet{reese2006b}.
They developed a two-dimensional spectral numerical approach to compute the acoustic modes in centrifugally distorted polytropes, including the full effect of the Coriolis force. 
According to their results, complete calculations are required for fast rotators with $v > 50$ km sec$^{-1}$.
The primary star, J0247-25A, is the fast rotator with a projected velocity of $v \sin i = 95 \pm 5 $ km sec$^{-1}$ (Maxted13).

\citet{reese2006a} presented a list of pulsation frequencies obtained from the complete calculation in Appendix F.2 of his doctoral thesis.
These frequencies were given in the co-rotating frame and must be transformed into the inertial frame by adding $-m \Omega$ to compare them with the observed frequencies.
Figure \ref{fig_rot_shift} displays some of these data as examples, all of which are expressed in the same unit as $\sqrt{(GM/R_{\tt pol}^3)}$.
For our stellar model, the polar radius, $R_{\tt pol}$, was deduced from $r_{\tt pole}$, listed in Table \ref{tab_binary}.

As shown in the upper panel of Figure \ref{fig_rot_shift}, the pulsation angular frequency changes smoothly with the increase in the rotation rate.
Therefore, the pulsation frequency for a given rotation rate can be estimated by linearly interpolating these data, and the frequency shift can then be calculated;
the pulsation frequency at the rotation rate $\Omega$ = 0 is defined as $\omega_0$ for each mode.
Since the frequency shift varies smoothly with the increase of $\omega_0$ along with the radial order, $n$ (lower panel of Figure \ref{fig_rot_shift}), 
these values were interpolated linearly to estimate the frequency shift for the given $\omega_0$ obtained from the GYRE.

The approach to derive the frequency shift for a realistic stellar model by interpolating the values from the polytropic model
is conceptually identical to those of \citet{saio1981} and \citet{perezhernandez1995}, who interpolated perturbative coefficients such as $C_1$ and $Z$.

\subsection{Grid-based Fitting}
For a given stellar model, the rotationally shifted frequencies were obtained from the above process, and these theoretical frequencies were fitted with the observed frequencies.
The fitting parameter, $\chi^2$, is defined as:
\begin{equation}
\chi^2 = \frac{1}{N} \sum_{i=1}^{N} \frac{(f_{theo,i} - f_{obs,i})^2}{\sigma_i^2},
\end{equation}
where the uncertainties, $\sigma_i$, were assigned equally to be 0.1 day$^{-1}$, following \citet{murphy2021}.
$f_{theo,i}$ and $f_{obs,i}$ are the theoretical and observed frequencies, respectively, and these cyclic frequencies ($f$ = $\omega$/2$\pi$) are in the unit of day$^{-1}$.

Since there are many more theoretical frequencies with the pulsation modes ($n$, $\ell$, $m$) than observed ones,
a combination of pulsation modes is determined to obtain a minimum $\chi^2$.
At this point, only the $\ell + |m| = even$ modes were considered because the disc-integrated lights of the other $odd$ ones show significantly lower amplitudes
for the eclipsing binary stars \citep{mkrtichian2018}, assuming that both the rotation and pulsation axes align with the orbital axis.

Considering that the high $\ell$ modes are less visible because of the disc-averaging effect \citep{daszkiewicz2002, ballot2011},
a fit with $\ell$ = 0--2 was attempted initially, but no reliable solutions with $\chi^2 <$ 1.0 could be found, which is a criterion suggested by \citet{murphy2021}.
Therefore, the angular degree extended up to $\ell$ = 3, and three frequencies were identified as $\ell$ = 3 modes (see Table \ref{tab_seismic}).

\begin{figure*}
\center\includegraphics[scale=0.83]{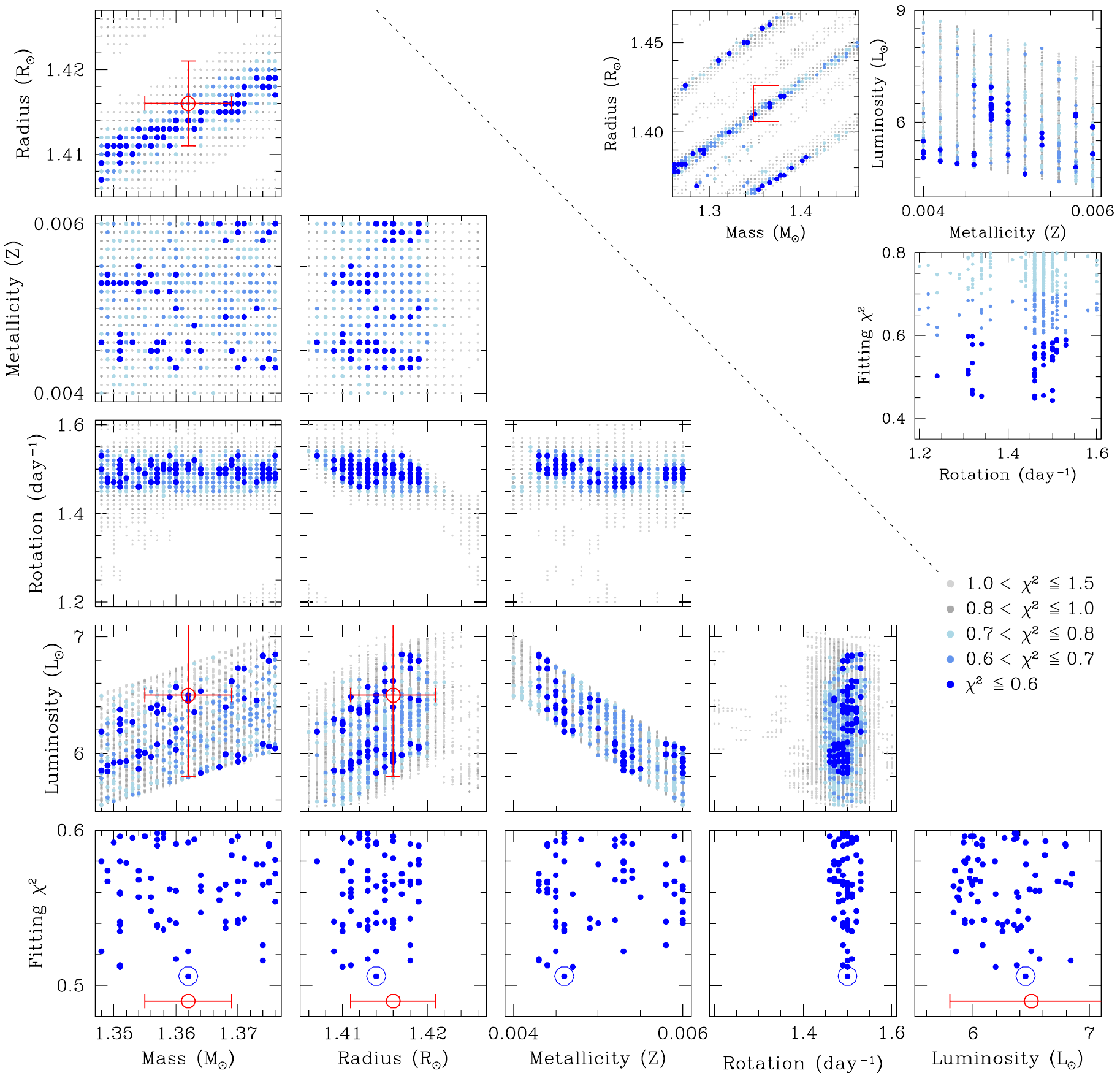}
\caption{Grid fitting plots. The AGSS09 composition and $\alpha_{\rm MLT}$ = 1.0, were used for the evolutionary models.
The three panels on the upper right are the results for a wide range of stellar masses and radii. The red box denotes two times the measured errors and
corresponds to the narrow one; the results are shown on the lower left side. The red circles with error bars represent the observed values of the mass, radius, and luminosity.
The blue dots with open circles in the bottom panels represent our adopted model. \label{fig_grid_fit}}
\end{figure*}

Grid-based fitting was conducted for various stellar models and rotation rates.
The stellar model grids were selected based on the physical parameters described in Section \ref{subsec_phy_par}.
Firstly, the wide range of masses and radii was examined from 1.262 to 1.462 $M_\sun$ with a step of 0.004 $M_\sun$ and from 1.366 to 1.466 $R_\sun$ with a step of 0.002 $R_\sun$, respectively.
The metallicities were determined from Z = 0.004 to 0.006 with a step of 0.0002. The rotation rates were taken from 1.2 to 1.6 day$^{-1}$ with a 0.02 day$^{-1}$ step,
considering that the rotation rate deduced from the projected velocity is 1.3 day$^{-1}$ and the synchronous rate with the binary orbit is 1.5 day$^{-1}$.
These steps of masses, radii, and rotation rates were determined to ensure a similar frequency difference of about 0.1 day$^{-1}$ between the steps, even differing from the pulsation mode to mode;
the metallicity step was relatively less sensitive.
As shown in the upper right panels of Figure \ref{fig_grid_fit}, several good fits with $\chi^2 <$ 0.8 spread out over nearly all the ranges.
The solutions were located along the same density lines, without a definite global minimum.

Therefore, the range of masses and radii was tightened as the second stage, considering the fact that
the observed mass and radius for the double-lined eclipsing binary J0247-25 were measured directly by analyzing both the photometric and spectroscopic data.
The narrow range was adopted to be twice that of the measured errors, i.e., masses from 1.348 to 1.376 $M_\sun$ with a step of 0.001 $M_\sun$ 
and radii from 1.406 to 1.426 $R_\sun$ with a step of 0.001 $R_\sun$. 
The ranges of metallicities and rotation rates remained the same, but their steps were reduced to 0.0001 in Z and 0.01 day$^{-1}$, respectively.
The lower left panels of Figure \ref{fig_grid_fit} present the detailed results for the narrow range.
Once again, there are several solutions along the same density line, and a global minimum is not visible.
The number of observed frequencies may be insufficient to make a good constraint on the stellar model parameters such as mass, radius, and metallicity. 
Nonetheless, it is impressive that the rotation rates are distributed very tightly around 1.50 day$^{-1}$, indicating synchronous rotation.

\begin{figure}
\center\includegraphics[scale=0.85]{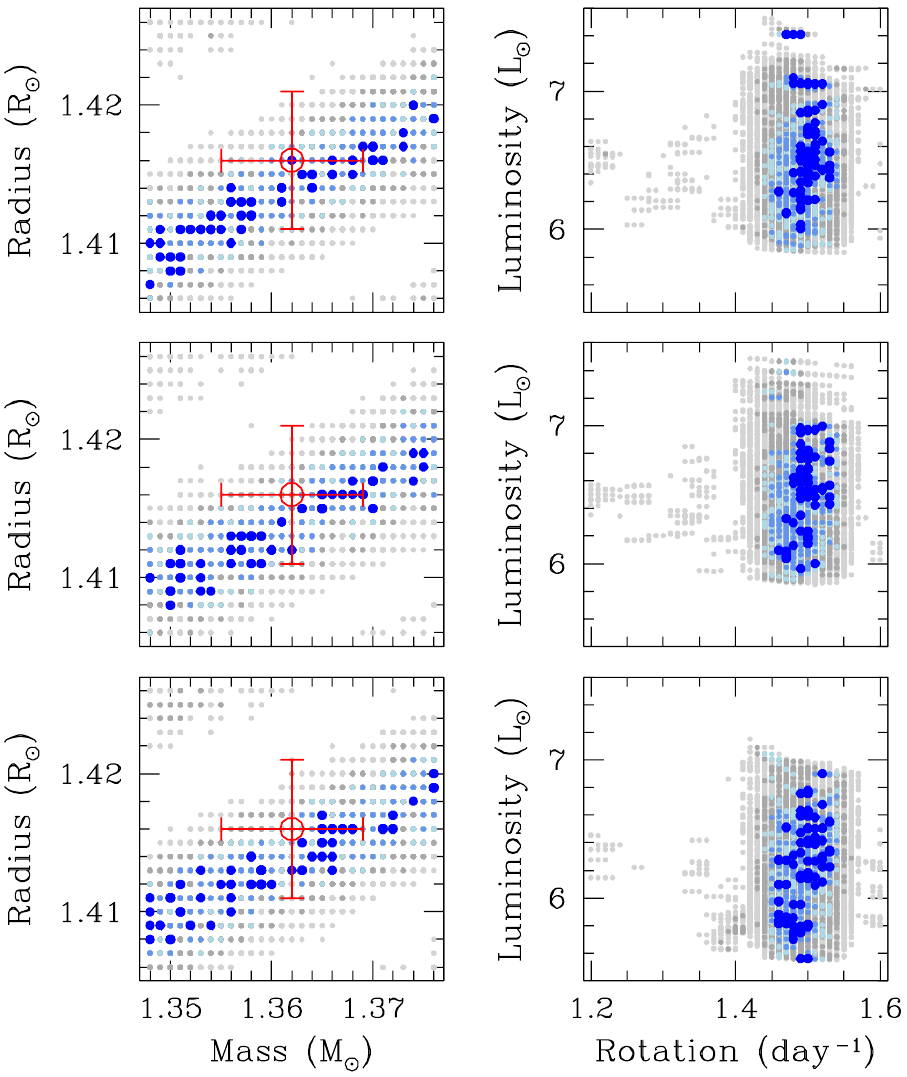}
\caption{Comparison of modellings with different parameters. The color codes and error bars are identical to those shown in Figure \ref{fig_grid_fit}.
(Top) With the GS98 composition and $\alpha_{\rm MLT}$ = 1.0, (middle) with the GS98 composition and $\alpha_{\rm MLT}$ = 1.8, 
(bottom) with the AGSS09 composition and $\alpha_{\rm MLT}$ = 1.8. \label{fig_diff_par}}
\end{figure}

The dependency on the evolutionary model parameters was analyzed.
The initial metal fraction and mixing length parameter were changed to be GS98 \citep{grevesse1998} and $\alpha_{\rm MLT}$ = 1.8, respectively.
Figure \ref{fig_diff_par} illustrates the results, which are very similar to those obtained using the parameters of AGSS09 \citep{asplund2009} and $\alpha_{\rm MLT}$ = 1.0,
as shown in Figure \ref{fig_grid_fit}.
The GS98 composition appears to produce higher luminosities than AGSS09, and the dependency on $\alpha_{\rm MLT}$ is not conspicuous.
Regardless of the model parameters, the rotation rates are sufficiently constrained to approximately 1.50 day$^{-1}$.

\subsection{The Best Solution \label{sec_solution}}
Consequently, a model with the lowest $\chi^2$ was selected as the best solution, 
of which the parameters with 1.362 $M_\sun$, 1.414 $R_\sun$, and 6.45 $L_\sun$ concur well with the observations, as shown in the bottom panels of Figure \ref{fig_grid_fit}.
The best solution showed Z = 0.0046, and its rotation rate was 1.50 day$^{-1}$.
These values coincide with the averages of 73 models with $\chi^2 \le$ 0.6: 1.362 $\pm$ 0.009 $M_\sun$, 1.414 $\pm$ 0.003 $R_\sun$, 6.25 $\pm$ 0.30 $L_\sun$, 
Z = 0.0050 $\pm$ 0.0006, and 1.495 $\pm$ 0.017 day$^{-1}$, in order.\footnote{The current results were obtained by using the 10 frequencies listed in Table \ref{tab_seismic}.
We confirmed that these results are approximately identical to those using eight frequencies,
excluding the two frequencies of $f_6$ and $f_{10}$ which were initially identified as combination frequencies;
for example, 1.364 $\pm$ 0.009 $M_\sun$, 1.416 $\pm$ 0.003 $R_\sun$, and 1.474 $\pm$ 0.022 day$^{-1}$. }

\begin{deluxetable*}{rccccc}
\tablewidth{0pt} \tablecolumns{6}
\tablecaption{Comparison of the observed and theoretical frequencies for the best solution. All frequencies have the same unit of day$^{-1}$. \label{tab_seismic}}
\tablehead{ \colhead{Observed $f$} & \colhead{ Pulsation Mode} & \colhead{Model $f_0$} &  \colhead{Shift $\delta f$} & \colhead{Model $f_{n,\ell,m}$} & Difference} 
\startdata
    $f_1$ = 34.176 & $(3, 2, +2)_{100}$  & 39.063 & $-4.872$ & 34.191 & $-0.015$ \\
    $f_2$ = 41.346 & $(3, 3, +1)_{56}$  & 44.583 & $-3.222$ & 41.361 & $-0.015$ \\
    $f_3$ = 43.713 & $(4, 2, 0)_{100}$    & 45.729 & $-2.043$ & 43.686 & $+0.027$ \\
    $f_4$ = 39.195 & $(4, 1, +1)_{58}$  & 42.701 & $-3.394$ & 39.307 & $-0.112$ \\
    $f_5$ = 52.120 & $(6, 0, 0)_{100}$    & 53.474 & $-1.379$ & 52.095 & $+0.025$ \\ 
    $f_6$ = 35.669 & $(3, 1, -1)_{99}$   & 35.842 & $-0.187$ & 35.655 & $+0.014$ \\       
    $f_7$ = 47.117 & $(5, 2, +2)_{100}$  & 52.716 & $-5.711$  & 47.005 & $+0.112$ \\
    $f_9$ = 40.402 & $(4, 2, +2)_{100}$  & 45.729 & $-5.300$ & 40.429  & $-0.027$ \\ 
$f_{10}$ = 36.862 & $(1, 3, -3)_{81}$   & 34.496 & $+2.293$ & 36.789 & $+0.073$ \\
$f_{11}$ = 47.481 & $(4, 3, -1)_{88}$   & 48.144 & $-0.531$ & 47.613 & $-0.132$ \\  
\enddata
\end{deluxetable*}

Table \ref{tab_seismic} lists the theoretical frequencies of the best solution, including the observed ones, for comparison.
The second column presents the theoretical pulsation modes $(n, \ell, m)_p$, where $p$ is the distribution rate (\%) among the models with $\chi^2 \le$ 0.6.
This shows that all the adopted modes hold the majority of the mode distribution.
The other modes for $f_2$ are $(2,3,-3)_{24}$ and $(4,3,+3)_{19}$, and those for $f_4$ are $(2,3,-1)_{23}$ and $(3,3,+1)_{19}$.
In the following columns, the theoretical frequencies $f_0$ for the non-rotating model were obtained using the GYRE,
and the frequency shifts $\delta f$ were derived by interpolating the complete calculation results.
The rotationally shifted theoretical frequencies $f_{n,\ell,m}$ were calculated by adding $f_0$ and $\delta f$.
The last column shows the difference between the observed and theoretical frequencies $f_{n,\ell,m}$.
The $\chi^2$ value of 0.506 is comparable with the results of \citet{murphy2021},
who found the best model with $\chi^2$ = 0.386 for the $\delta$ Sct-type pulsator HD139614. 

\begin{figure}
\center\includegraphics[scale=0.52]{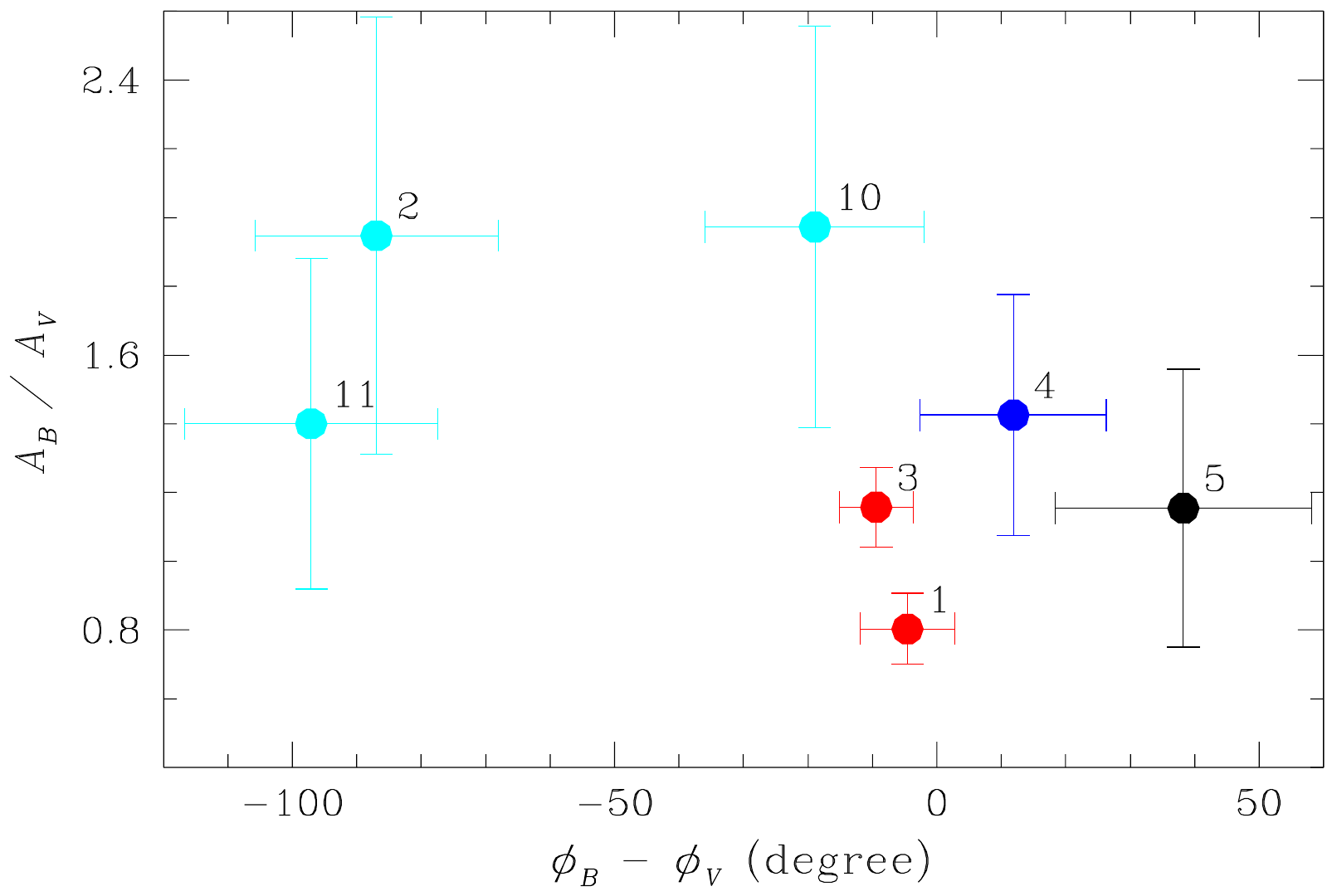}
\caption{Phase differences and amplitude ratios between two passbands, $B$ and $V$. The filled circles with error bars represent the observed values of seven frequencies, numbering $f_i$.
The colors denote the theoretical pulsation modes in Table \ref{tab_seismic}: black for $\ell$ = 0, blue for $\ell$ = 1, red for $\ell$ = 2, and cyan for $\ell$ = 3. \label{fig_mode}}
\end{figure}

The angular degree $\ell$ can be determined from the amplitude ratio and phase difference between the passbands \citep{watson1988, garrido1990}.
The amplitude ratios, $A_B / A_V$, and the phase differences, $\phi_B - \phi_V$, were obtained for 12 {\it TESS} frequencies by fitting these frequencies to the KMTNet $B$ and $V$ data.
Figure \ref{fig_mode} displays the results of the seven frequencies because the others contain considerably large errors.
This shows that the phase difference decreases with the increase in the $\ell$ value and that the $\ell$ = 3 mode has a high amplitude ratio, although the observation errors are not small.
These tendencies are consistent with the theoretical predictions \citep{watson1988, moya2004} and other observation results \citep[for example,][]{breger2017, paparo2018}. 
Therefore, the theoretical pulsation modes identified in this study appear to be reliable.

\begin{figure}
\center\includegraphics[scale=0.64]{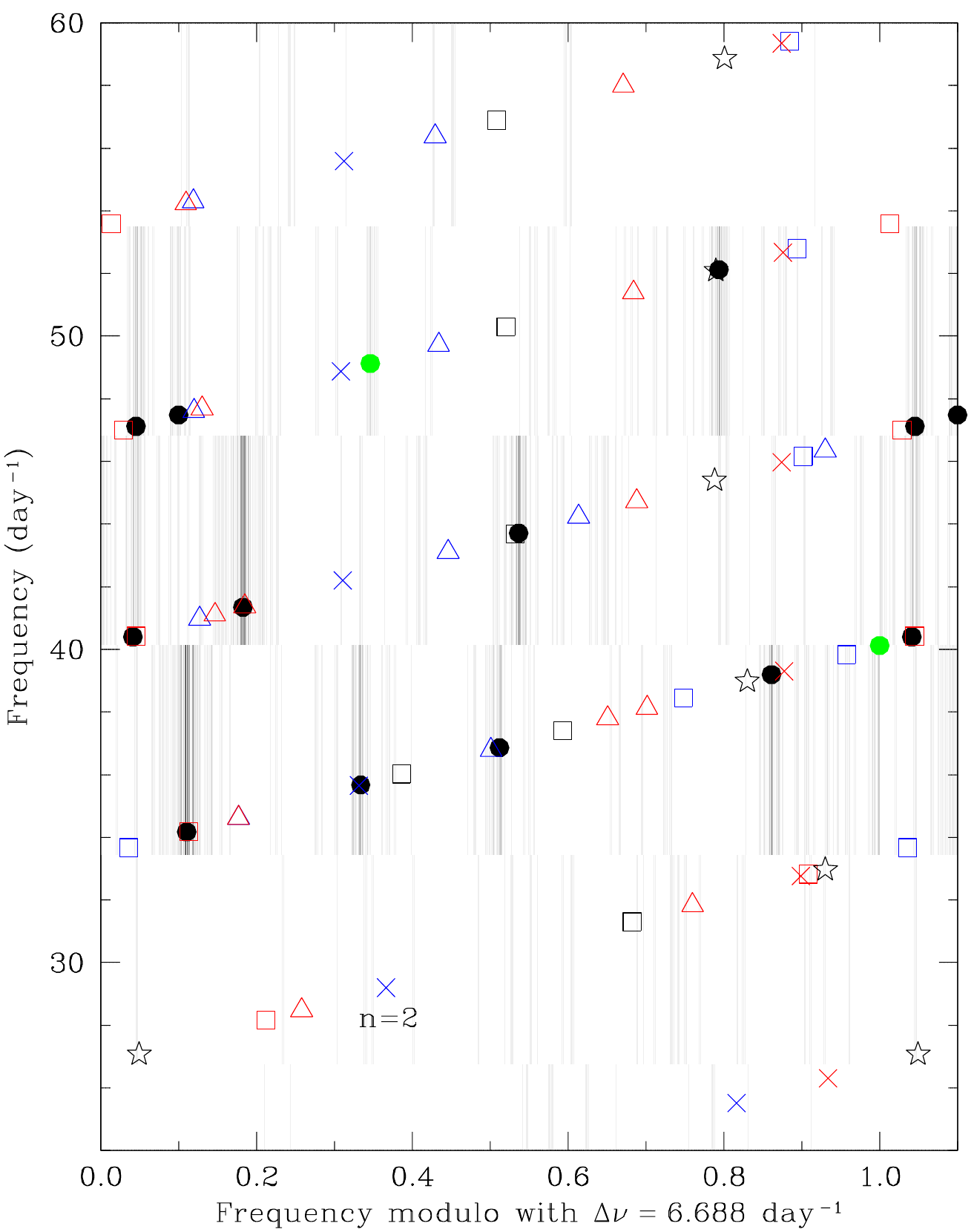}
\caption{\'Echelle diagram. The observed 12 frequencies are represented by filled circles,
of which the green colors represent the two combination frequencies ($f_8$ and $f_{12}$) unused in the model fitting. 
The star symbols, crosses, rectangles, and triangles represent the theoretical frequencies, $f_{n,\ell,m}$, with $\ell$ = 0, 1, 2, and 3, respectively.
Their colors indicate the motion: blue for prograde modes, black for zonal modes,  and red for retrograde modes.
The gray-scale vertical bars represent the Fourier amplitude of the {\it TESS} data. \label{fig_echelle}}
\end{figure}

Figure \ref{fig_echelle} presents the \'echelle diagram for the best solution, showing that the observed frequencies concur well with the theoretical ones. 
The large frequency separation, $\Delta \nu$ = 6.688 day$^{-1}$, was derived by averaging the theoretical frequencies with $n$ = 4--6 and $\ell$ = 1 modes.
The separation value is very close to the difference of 6.715 day$^{-1}$ between the observed frequencies, $f_7$ and $f_9$, with identical $\ell$ = 2 and $m$ = +2 mode,
which are vertically aligned in the diagram.
It also concurs well with 6.7 $\pm$ 0.3 day$^{-1}$, deduced from the empirical relation by \citet{garciahernandez2017}, using the observed density of 0.480 $\pm$ 0.008 $\rho_\sun$ for J0247-25A.

\section{Discussion and Conclusion \label{sec_discuss}}
\subsection{Rotation of the Binary Component}
Most eclipsing binaries with ellipsoidal variations are likely to be circularized and synchronized \citep{lurie2017},
even if the ellipsoidal variation caused by the deformation of the star's shape is not a direct measurement of the stellar rotation.
The asynchronous rotation of these binaries appears to be uncommon but is not unprecedented.
For example, \citet{lwang2020} presented the spectroscopic results of the prototype star, EL CVn, which showed well-defined ellipsoidal variations \citep{maxted2014}.
They estimated the projected rotation velocity of the A-type primary star to be $v \sin i$ = 64.8 $\pm$ 1.0 km sec$^{-1}$, which is much slower than the synchronized velocity of 91 km sec$^{-1}$,
whereas the rotation velocity of the pre-He-WD secondary was coincident with the synchronous value.

For the primary star of our target, J0247-25, Maxted13 measured the projected equatorial rotation velocity to be $v_{\rm eq} \sin i$ = 95 $\pm$ 5 km sec$^{-1}$ 
from the Doppler broadening of the spectral lines.
This corresponds to 87\% of the synchronized velocity of 109 $\pm$ 1 km sec$^{-1}$, which is calculated with the orbital period and the equator radius, $R_{\rm eq}$ = 1.433 $\pm$ 0.008 $R_\sun$, 
averaging the side and back radii in Table \ref{tab_binary}. The orbital inclination, $i$, is used, assuming that the rotation axis aligns with the orbital axis.
Despite the presence of some uncertainties at $v_{\rm eq} \sin i$ and $R_{\rm eq}$, the measured velocity favors the asynchronous rotation of the primary star.

Conversely, our seismic analysis constrained the rotation rate to 1.50 $\pm$ 0.02 day$^{-1}$, perfectly concurring with the synchronous rotation rate of 1.4974 day$^{-1}$.
There was no good solution with $\chi^2 \le$ 0.6 found near the rotation rate of 1.33 day$^{-1}$ determined from the measured velocity.
Therefore, it can be inferred that the primary star, J0247-25A, rotates synchronously with the binary orbit.

As described in Section \ref{sec_rotshift}, the seismic analysis uses the polar radius to derive the rotationally shifted frequencies.
The polar radius was obtained from the radius value of our stellar model grids multiplied by the ratio of pole to volume radius (that is, $r_{\tt pole} / r_{\tt volume}$, listed in Table \ref{tab_binary}).
The ratio is 1.0 for a non-rotating spherical star and decreases with the increase of the rotation rate.
For example, the ratio was estimated to be 0.85 $\pm$ 0.01 for the rapidly rotating star, Altair, with the angular velocity of $\Omega / \Omega_K$ = 0.744,
based on the interferometric data and modeling \citep{bouchaud2020}. Here, $\Omega_K$ is the Keplerian angular velocity at equator.
J0247-25A with $\Omega / \Omega_K$ = 0.224 is presumed to have a ratio higher than 0.95.
In this study, the ratio value of 0.976 $\pm$ 0.006, deduced from the binary model, was used.
We evaluated the uncertainty due to the polar radius by changing the ratio of 0.976 to 1.0 and 0.95.
Other stellar properties were consistent in all three cases within the error range,
whereas the rotation rate appeared to increase slightly with decreasing the ratio (or decreasing the polar radius):
1.457 $\pm$ 0.024 day$^{-1}$ for the ratio of 1.0, 1.495 $\pm$ 0.017 day$^{-1}$ for the ratio of 0.976, and 1.536 $\pm$ 0.022 day$^{-1}$ for the ratio of 0.95.
The ratios of rotation to orbital rate are very close to 1.0 (0.97--1.03), indicating a synchronous rotation \citep{lurie2017}.

\subsection{Seismic Analysis of Fast-rotating Pulsators}
$\delta$ Sct-type pulsating stars with multiple frequencies have long been considered as an important object in asteroseismology \citep{brown1994}. 
Several multisite observation campaigns have been conducted to obtain their frequencies with the highest possible accuracy \citep[for a review]{breger2000},
and many stars have been analyzed using the space-based continuous data \citep[for example,][]{balona2015, paparo2016, bedding2020}. 
However, asteroseismology of the $\delta$ Sct stars has not yet been fruitful, mainly owing to the difficulty of mode identification \citep{goupil2005, cunha2007, handler2013}.
In the $\delta$ Sct stars, the distribution of mode amplitudes is very irregular, and not all modes are excited to observable amplitudes.
Their fast rotation deteriorates the situation because it destroys the equidistance of the rotational $m$-mode splitting, resulting in irregular frequency spacing \citep{goupil2000}.
Recently, a few seismic interpretations of fast-rotating $\delta$ Sct stars have been performed using realistic two-dimensional stellar models;
for example, \citet{zwintz2019} for $\beta$ Pic and \citet{bouchaud2020} for Altair.

In this study, a seismic analysis of the $\delta$ Sct-type pulsating star J0247-25A was conducted using the observed 10 frequencies.
Theoretical frequencies were obtained by adding the non-rotating model frequencies from the GYRE and their rotational shifts from the complete calculation approach.
The best solution with the lowest $\chi^2$ was determined from the grid-based fitting for various stellar properties.
Its theoretical frequencies and stellar parameters (mass, radius, and luminosity) concurred well with the observations.
Furthermore, the observed values of the phase differences and amplitude ratios clearly demonstrated that the pulsation modes identified theoretically, are reliable.

The results imply that the complete calculation approach based on the polytropic model \citep{reese2006b} is applicable to the analysis of a real star.
To our knowledge, J0247-25A is the first object that the complete approach is applied to the fast-rotating $\delta$ Sct star.
Accordingly, further objects will be required to validate this application.
We will focus on double-lined eclipsing binaries with $\delta$ Sct-type pulsating components
because they can provide a valuable constraint on the stellar properties
and present an advantage in the mode identification framework as only the $\ell + |m| = even$ modes are visible.
In this attempt, several new pulsation frequencies of the eclipsing binary AB Cas were detected with the {\it TESS} data,
and satisfactory results were obtained ({\it in preparation}).

\begin{acknowledgements}
SLK wishes to thank Dr. D. Reese for his kind explanations of the complete calculation and valuable comments on this manuscript.
This research has made use of the KMTNet system operated by the Korea Astronomy and Space Science Institute (KASI) and
the data were obtained at two host sites of SAAO in South Africa and SSO in Australia.
This paper includes data collected by the {\it TESS} mission. Funding for the {\it TESS} mission is provided by the NASA's Science Mission Directorate.
This research was supported by the KASI grant 2021-1-830-08.
KH was supported by the grants 2017R1A4A1015178 from the National Research Foundation (NRF) of Korea.
We would like to thank Editage (www.editage.com) for English language editing.
\end{acknowledgements}

\facility{KMTNet \citep{kim2016a}, TESS \citep{ricker2015}}


\end{document}